\documentclass[sigconf]{acmart}

\AtBeginDocument{%
  }

\setcopyright{cc}
\setcctype{by-nc-nd}
\copyrightyear{2026}
\acmYear{2026}
\acmDOI{10.1145/3772363.3799129}
\acmConference[CHI EA '26]{Extended Abstracts of the 2026 CHI Conference on Human Factors in Computing Systems}{April 13--17, 2026}{Barcelona, Spain}
\acmBooktitle{Extended Abstracts of the 2026 CHI Conference on Human Factors in Computing Systems (CHI EA '26), April 13--17, 2026, Barcelona, Spain}
\acmISBN{979-8-4007-2281-3/2026/04}

\begin{document}

\title{\emph{Magical Touch}: Transforming Raw Capacitive Streams into Expressive Hand-Touchscreen Interaction}

\author{Yuanlei Guo}
\email{yguo623@gatech.edu}
\affiliation{%
  \institution{Georgia Institute of Technology}
  \city{Atlanta}
  \country{USA}
}

\author{Xizi Gong}
\email{gongxizi0730@pku.edu.cn}
\affiliation{%
  \institution{Microsoft Research Asia}
  \city{Beijing}
  \country{China}
}

\author{Yizhong Zhang}
\email{yizhong.zhang@microsoft.com}
\affiliation{%
  \institution{Microsoft Research Asia}
  \city{Beijing}
  \country{China}
}

\author{Xiaoyu Zhang }
\authornote{Corresponding Author}
\email{xiaoyu.zhang@cityu.edu.hk}
\affiliation{%
  \institution{City University of Hong Kong}
  \country{Hong Kong SAR}
}

\begin{abstract}
Modern touchscreens utilize capacitive sensing technology to enable precise and robust multi-touch interaction. However, the broader expressive potential of the human hand remains underutilized, since most existing methods directly filter out larger-area hand–screen contact. This paper introduces \emph{Magical Touch}, an interaction method based on raw capacitive sensing data. By directly integrating raw touchscreen sensor data into the interaction loop, our method allows users to interact with the screen naturally and efficiently using arbitrary hand gestures on existing touchscreen devices. To demonstrate the feasibility and expressive capacity of this approach, we implement a physics-based interactive game featuring single-player, multiplayer collaborative, and pressure-sensitive modes. These scenarios showcase how digital objects can respond in real-time to both the geometry and contact intensity of the user's hand. Our results indicate that leveraging raw capacitive data can expand the design space of touchscreen interaction, offering an embodied and continuous interaction paradigm beyond existing fingertip-based approaches.
\end{abstract}



\maketitle
\renewcommand{\shortauthors}{Yuanlei Guo et al.}

\section{Introduction}

As smartphones and tablets have become the primary electronic devices in everyday life, touchscreen interaction has emerged as the most important form of human–computer interaction. Touchscreen technology has evolved from early resistive screens that supported only single-point input~\cite{hurst1970touch} to mutual-capacitive displays enabling multi-touch interaction~\cite{dietz2001diamondtouch,rekimoto2002smartskin,boie1984capacitive}, and has become the de facto industry standard following its widespread adoption in consumer devices since the first release of iPhone in 2007~\cite{orphanides2017touchscreen}. As a window through which humans engage with the digital world, display sizes have continued to grow, from the 3.5-inch screen of the first-generation iPhone to the 6.9-inch display of the iPhone 17 Pro Max. More recently, tri-fold devices (e.g., Huawei Mate XT and Galaxy Z TriFold) offer unfolded screen sizes exceeding 10 inches. These changes in screen size and form factor introduce new challenges and opportunities for touchscreen interaction.

\begin{figure*}
  \centering
  	\includegraphics[width=.95\textwidth]{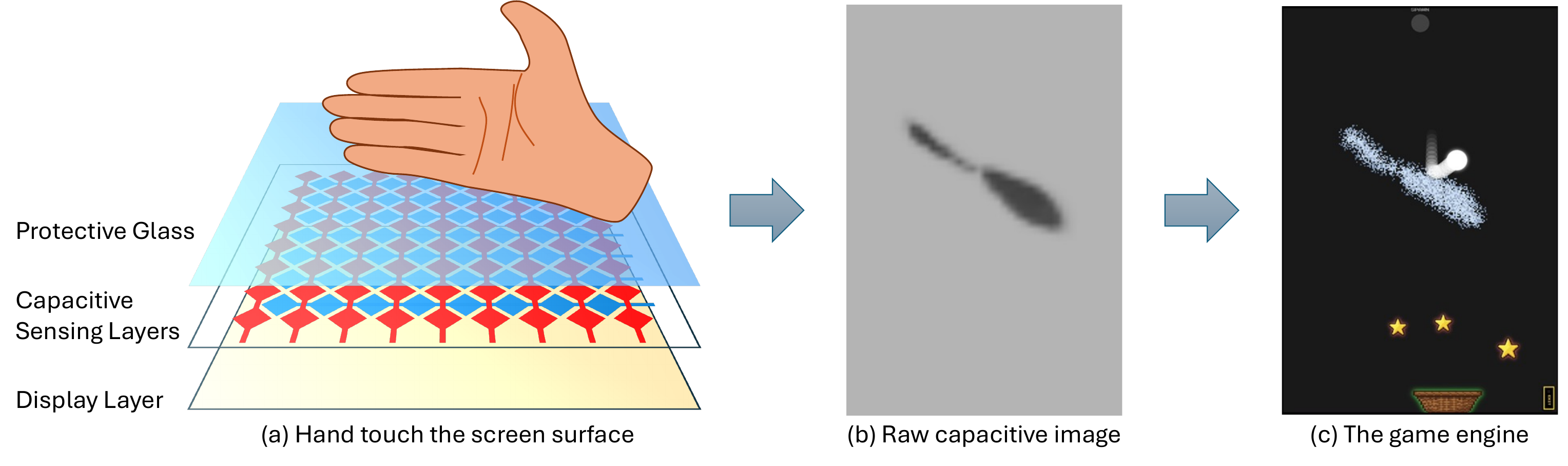}
  \Description{A three-step pipeline diagram showing the proposed interaction method. Step (a) shows a hand touching a capacitive touchscreen surface. Step (b) shows the resulting two-dimensional raw capacitance image as a heatmap reflecting contact shape and force. Step (c) shows this capacitance image being fed into the game engine for natural and continuous interaction.}
  \caption{Overview of the proposed interaction method based on raw capacitance data.
  (a) When a hand touches the surface of a capacitive touchscreen, the sensor detects changes in the spatial capacitance distribution, thus obtaining
  (b) a two-dimensional raw capacitance image. This image reflects the overall contact shape and force between the hand and the screen. 
  (c) This raw capacitance image is directly input into the interaction system (the game engine) to achieve natural and continuous interaction. 
  }
  \label{fig:method_overview}
\end{figure*}

Despite the increasing diversity of touchscreen devices in terms of size, form factor, and application scenarios, their underlying sensing principles remain largely unchanged. Most capacitive touchscreens rely on mutual-capacitance scanning to capture charge distributions induced by skin contact, identifying prominent local peaks as fingertip touches while suppressing other regions as noise~\cite{westerman1999hand}. This approach enables precise and robust detection of fingertip locations for reliable interaction, but it also constrains the expressiveness of supported gestures. For example, when a user places their palm on the screen, large and continuous contact regions are often filtered out by touch detection algorithms and thus cannot be effectively recognized. At the same time, transient local peaks generated at the moment of palm contact may be mistakenly interpreted as fingertip touches, leading to unintended inputs~\cite{mohamed2014efficient}. For large-screen devices such as tablets and touchscreen laptops, users are more likely to adopt diverse hand postures during interaction~\cite{yang2025interaction}. Enabling interaction modalities beyond fingertip-based input therefore holds the potential to improve both the efficiency and naturalness of interaction on large touch surfaces.

To overcome the limitations of conventional fingertip-based touchscreen interaction and enable more natural forms of hand interaction, recent research has begun to directly leverage raw sensing data from touchscreens to achieve finer-grained modeling of hand contact. Such work reads the raw two-dimensional capacitive sensing data acquired by touchscreens and applies geometric analysis or machine learning methods to classify gestures~\cite{huang2024specifingers, le2018palmtouch} or estimate hand poses~\cite{ahuja2021touchpose, liu2024touchscreen, DBLP:conf/mc/SchweigertLHWLM19, choi20213d}, thereby supporting interaction techniques that go beyond traditional touch-point–level input. Notably, capacitive touchscreens are not limited to sensing the human hand itself; in principle, any object capable of inducing capacitive changes can be detected and modeled. Accordingly, some studies exploit raw capacitive sensing data to recognize tags attached to small objects and combine them with on-screen visual content, enabling interactive applications centered on tangible object manipulation~\cite{steuerlein2022conductive,rusu2023deep,schmitz2021itsy}.

This paper follows the same research direction by using raw touchscreen sensing data as interaction input; however, unlike prior work, we do not attempt to explicitly recognize or estimate specific gestures or hand poses. Instead, we directly integrate the raw sensing data into the interaction process itself, allowing the overall contact configuration between the hand and the screen to directly drive interaction behavior, thereby avoiding the complexity introduced by gesture definition and recognition. Guided by this design perspective, the focus of this paper is not to propose a new gesture recognition algorithm, but to explore a touchscreen interaction paradigm centered on raw sensing data that complements traditional fingertip-based touch, and to demonstrate its feasibility and expressive potential in practical application scenarios through concrete interaction examples. 

To demonstrate the interactivity offered by this approach, we implemented a simple physics-based game. In this game, a series of balls fall under gravity and are directly controlled by the user's hand contact. All areas of contact with the screen are uniformly treated as obstacles affecting the ball's movement, which allows the user to interact through natural hand gestures and contact force, without relying on predefined gestures or explicit gesture recognition.

\section{Design Strategy and Prototype}

\begin{figure}
  \centering
  	\includegraphics[width=\columnwidth]{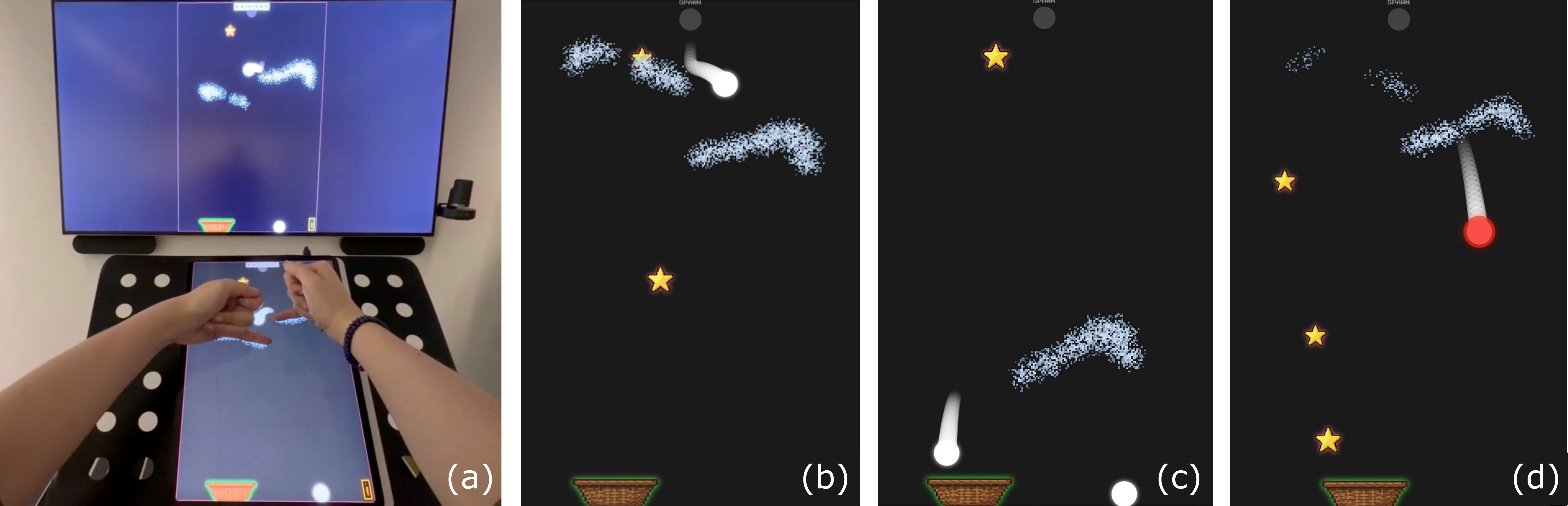}
  \Description{A four-panel figure showing single-player mode gameplay. Panel (a) shows a player touching the touchscreen with their hand. Panel (b) shows the hand contact area rendered as a gravel surface on screen. Panel (c) shows the ball being guided into a target basket. Panel (d) shows a ball turning red and exploding after falling freely for too long.}
  \caption{Single-player mode demonstration. (a) When the player touches the screen, the corresponding contact area is rendered as a \textit{gravel} surface as shown in (b). Players need to continuously create new surfaces to bounce the ball and collect stars along the way. The ultimate goal is to (c) guide the balls into the target basket. (d) The magic balls have a fixed lifetime and will turn red and explode if they fall freely for too long without bouncing.}
  \label{fig:single_player}
\end{figure}

To visually demonstrate the natural hand interaction capabilities based on raw touchscreen sensing data, we designed an interactive game centered on the physical movement of balls, revolving around the concept of "direct contact and manipulation of virtual objects with the hand". In the game, a series of balls continuously fall from the top of the screen under the influence of gravity. As illustrated in Fig.~\ref{fig:single_player}, users do not need to use specific gestures or control interfaces; instead, they place their hands directly on the screen as virtual obstacles, interacting with or colliding with the virtual balls to alter their trajectories and ultimately guide them to target locations. The entire interaction process emphasizes immediacy and continuity, allowing users to interact directly with virtual objects on the screen as if catching, lifting, or bouncing objects in the real world, thus reducing the learning cost and highlighting the natural interaction characteristics enabled by raw sensing data.

The game uses a physics engine to model the motion of the balls, including basic physical behaviors such as gravity, collision, and rebound. Building upon this foundation, the system further incorporates changes in contact area reflected in the raw touchscreen sensing data into the physical calculations to approximate the pressure applied by the hand to the screen: when a user presses the screen more firmly, the contact area between the skin and the touchscreen increases, resulting in a stronger rebound effect in the game. This design makes the motion of the balls sensitive not only to the position and shape of the hand, but also to the magnitude of the applied force. Because these behaviors align with users' intuitive expectations of the physical world, the motion of the balls is highly predictable. Users can anticipate ball trajectories based on their intuitive understanding of gravity and rebound, and then perform fine-grained control by adjusting hand posture and applied pressure. Compared to control methods based on discrete gesture, this continuous physical feedback further strengthens the direct mapping between hand movements and screen responses, making the interaction process more natural and smooth.

In terms of specific game mechanics, we assign each ball a finite lifespan and define a clear target area, introducing scoring and failure conditions to improve interaction tempo and enhance task orientation. The game supports both single-player and two-player cooperative modes. In the latter, balls can be continuously transferred between two screens, requiring real-time collaboration between users. While these designs enhance game efficiency and challenge, they also place higher demands on interaction response speed, making it difficult for users to rely on complex or high-latency control strategies. Instead, users are encouraged to adopt fast and intuitive hand contact for control, further highlighting the advantages of natural hand-based interaction in dynamic tasks.

\subsection{Demo 1: Single-player Ball Rescue}

In the single-player mode, the technology bridges physical inputs with digital physics by mapping real-time hand contact to a transient terrain called \textit{"gravel"}. When the player places their hand on the screen, the system captures the raw contact area and instantly translates it into a physical field within the game engine. The ball's motion follows Newtonian dynamics, with its state updated each frame by the vector sum of gravity, velocity damping (linear drag), and a repulsive force generated by the gravel field. This repulsive force allows the ball to bounce off the user-generated surface as if it were a solid wall; specifically, the direction of the bounce is determined by the spatial gradient of the contact area, allowing the ball to deflect naturally off the slanted edges of the hand's silhouette. 

Capacitive sensor data value generated by player touch are mapped onto a discrete grid as gravel density values. Visually, this contact area is rendered as a gravel texture. Physically, it serves as the foundation for the collision model. The mapping is intensity-dependent: higher capacitive signal intensity (simulating firm pressure) creates denser, tightly packed gravel with a higher coefficient of restitution, while lower intensity results in a sparser distribution that dampens the ball's momentum. This density field is smoothed via a Gaussian kernel and expanded slightly to ensure fluid, stable interactions without collision glitches.

The player needs to continuously create new surfaces by moving their hands, reacting dynamically to the ball's bouncing path. The primary objective is to guide the randomly spawned balls into a target basket at the bottom of the screen, while ensuring the ball does not free-fall for an extended duration. This creates an engaging interactive loop of prediction, action, and adaptation, showcasing the expressive potential of raw capacitive sensing.

\begin{figure}
  \centering
  	\includegraphics[width=\columnwidth]{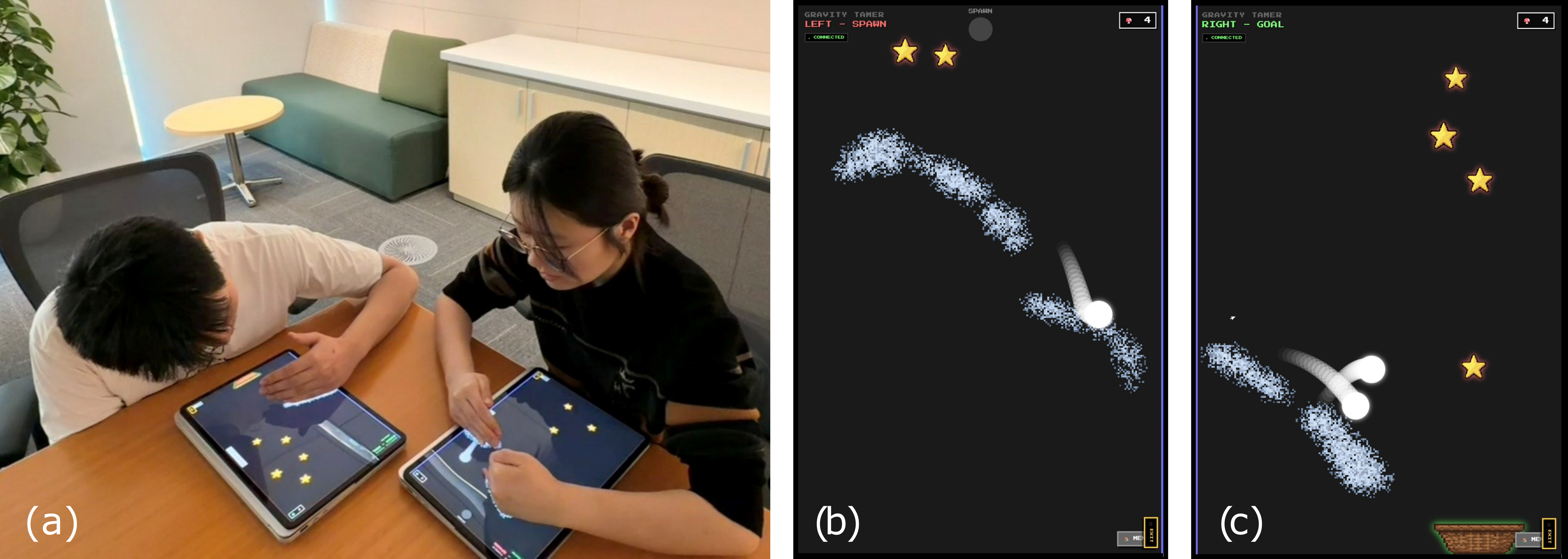}
  \Description{A three-panel figure showing multiplayer collaboration mode. Panel (a) shows two users sitting side by side, each interacting with their own touchscreen device. Panels (b) and (c) show the two screens forming a continuous virtual space where a ball traverses the boundary between screens, requiring coordinated hand placement to guide the ball into a target basket in the lower-right corner.}
  \caption{Multi-player collaboration mode demonstration.
  (a) Two users collaboratively play the game using touchscreen devices placed side by side.
  (b, c) The two players' screens jointly form a continuous virtual space. The ball can traverse the boundary between the two screens; therefore, the players must coordinate closely and respond rapidly to guide the ball into the target basket located in the lower-right corner.}
  \label{fig:multiple_player}
\end{figure}

\subsection{Demo 2: Multi-player Collaborative Mode}

The multi-player mode introduces a distributed collaboration mechanism where two players interact on two separate Microsoft Surface devices connected via a local network (Fig.~\ref{fig:multiple_player}). As the game environment spans across both tablets, players must perform cross-screen relays to guide the ball toward the target basket. This design requires constant collaboration and trajectory prediction, as players must proactively construct physical surfaces to receive or redirect the ball as it enters their respective zones.

\begin{figure*}
  \centering
  	\includegraphics[width=\textwidth]{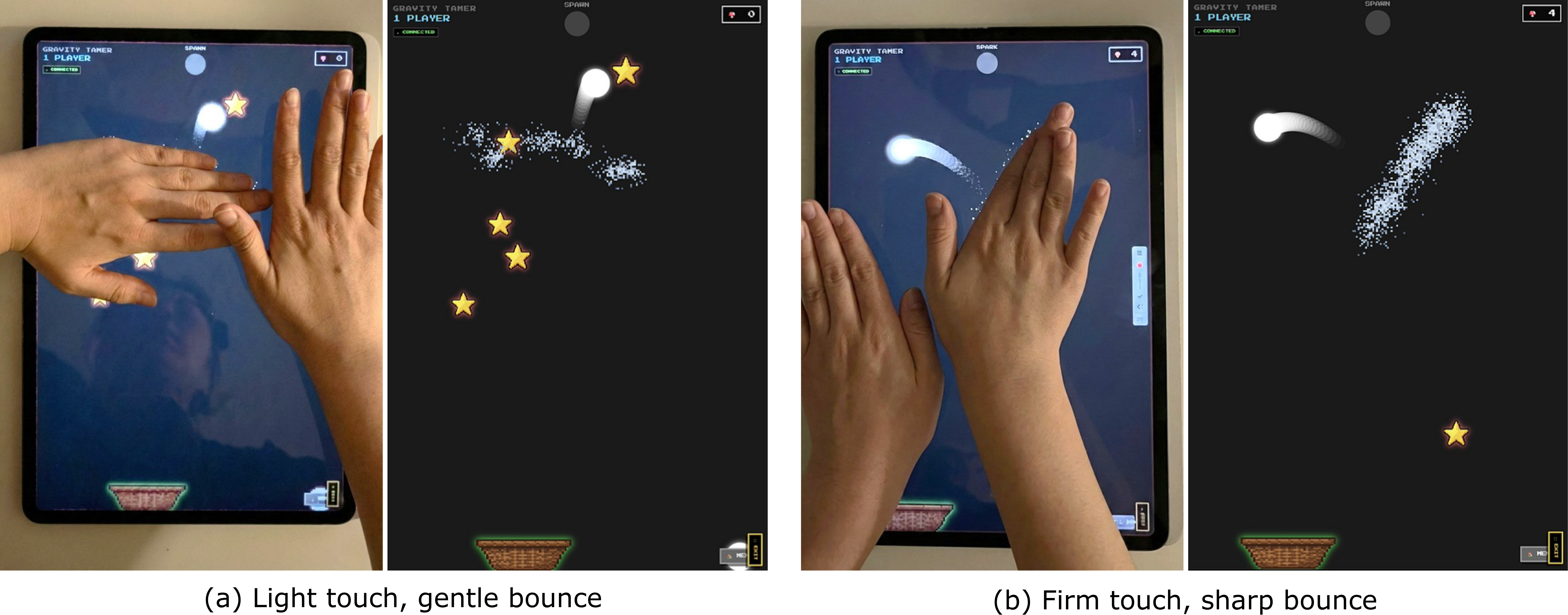}
  \Description{A two-panel comparison illustrating pressure-sensitive interaction. Panel (a) shows a light touch producing a sparse gravel surface that causes a gentle, low-angle ball deflection. Panel (b) shows a firm press producing a dense gravel surface that causes a sharp, high-velocity ball bounce.}
  \caption{An illustration of the pressure-sensing of \emph{Magical Touch}. 
  (a) Light touches generate sparse surfaces for gentle, low-angle deflections, while (b) firm presses create dense surfaces for sharp, high-velocity bounces.}
  \label{fig:pressure}
\end{figure*}

This setup fosters a highly dynamic and emergent gameplay experience. Since the hand-generated terrain is transient and its geometry shifts with the player's movements in real-time, the collaboration becomes a continuous process of mutual adaptation. For instance, if one player creates a steep slope at the screen edge to launch the ball at high speed, the partner must immediately adjust their hand's morphology to intercept or cushion the arrival. This transition from individual spatial control to real-time interpersonal coordination demonstrates \emph{Magical Touch}'s potential for complex, embodied social interaction.

\subsection{Demo 3: Pressure-sensitive Bouncing}

This demonstration illustrates how contact pressure directly modulates bounce dynamics. By analyzing raw capacitive intensity, the game engine maps hand applied force to the density of the \textit{gravel} field. This density-based visualization provides players with immediate visual feedback, which is then translated into tangible physical interactions. It creates a predictable physical relationship: light touches generate sparse surfaces for gentle, low-angle deflections, while firm presses create dense surfaces for sharp, high-velocity bounces. Fig.~\ref{fig:pressure} illustrates this mapping. This mechanism enables skill-based expressive control, allowing players to navigate complex trajectories through nuanced physical effort without additional force-sensing hardware. This provides an intuitive understanding of the pressure-to-bounce relationship and enables advanced, skill-based gameplay.

\section{Implementation Setup}

\paragraph{Hardware Configuration}
Our system runs on a Microsoft Surface Laptop Studio 2 device equipped with a 14.4-inch mutual-capacitive touchscreen. It is worth noting that commercially available Surface devices and their public system interfaces only provide fingertip touch data to developers by default, and do not support direct access to raw touchscreen data (i.e., the raw capacitive image). In this study, we collaborated with the Microsoft Surface team to provide a customized firmware update for the experimental device and obtained a system-level program written in C$\#$ to directly read the raw capacitance data from the operating system. This data is output as a two-dimensional matrix with a spatial resolution of $78\times52$ and a non-fixed sampling frequency, averaging approximately 100 Hz. The acquired raw data is then normalized and mapped to a numerical range of 0–255 for subsequent processing and visualization. This process demonstrates that, from a hardware and system implementation perspective, providing a raw sensing data interface for touchscreen devices does not present fundamental technical obstacles and is highly feasible for device manufacturers.

\paragraph{Software Implementation}
The game is implemented in JavaScript and runs in a web browser. Raw touchscreen perception data is acquired in real time by the underlying program and transmitted to the web browser, serving as continuous two-dimensional input for interactive calculations. The ball's movement is modeled based on a simple set of physics rules, including gravity and collisions and bounces with the hand's contact area. These rules are implemented using lightweight functions, designed to provide continuous and predictable motion behavior rather than precise physical simulation. The contact area between the hand and the screen is directly treated as a dynamic obstacle; its spatial distribution continuously affects the ball's movement, eliminating the need for gesture recognition or semantic parsing. In the two-player cooperative mode, the system synchronizes game state across multiple devices via WebSocket, enabling continuous ball movement across screens.

\section{Conclusion}

This paper introduces \emph{Magical Touch}, an interaction method based on raw capacitive touchscreen data, which demonstrates the potential of raw capacitive sensing for continuous hand-based interaction. As an extension to existing fingertip-based interaction paradigms, \emph{Magical Touch} directly leverages the geometry and signal intensity of the hand's contact with the screen, enabling a higher degree of freedom in human–computer interaction. We demonstrate the feasibility of raw data–driven interaction on existing capacitive touchscreen devices through a game featuring three scenarios: single-player interaction, multiplayer collaboration, and pressure-sensitive interaction. \emph{Magical Touch} opens up new possibilities for intuitive, flexible, and rapid touchscreen interactions. Future work will explore the integration of more advanced hand pose estimation techniques and extend this approach to collaborative productivity tools, further leveraging the full expressive capacity of the human hand on ubiquitous touchscreen devices.

\begin{acks}
We are grateful to Federico Zannier and Jonathan Westhues for their help in developing the firmware for capturing raw touch data on Microsoft Surface, and to Qiaochu Wang for his suggestions to improve the demo design.
We also thank the participants who volunteered for our pilot study and the anonymous reviewers for their feedback and suggestions that helped improve the quality of this paper.
This research is sponsored in part by the CityUHK Start-Up Grant (Project No. 9610708).
\end{acks}

\bibliographystyle{ACM-Reference-Format}
\bibliography{reference}

\end{document}